\documentclass[prd,showpacs,showkeys,preprintnumbers,aps]{revtex4}
\usepackage{graphicx}

\begin{document}
\draft

%
%

\preprint{TH-945}
\title{Quantum Hall States of Gluons  
in Quark Matter}
\author{Aiichi Iwazaki}
\address{Department of Physics, Nishogakusha University, Shonan Ohi Chiba
  277-8585,\ Japan.} 
\author{Osamu Morimatsu, Tetsuo Nishikawa}
\address{Institute of Particle and  Nuclear Studies, High Energy
  Accelerator Research Organization, 1-1, Ooho, Tsukuba, Ibaraki,
  305-0801, Japan}
\author{Munehisa Ohtani}
\address{Radiation Laboratory, RIKEN (The Institute 
of Physical and Chemical  Research), Wako, Saitama 351-0198, Japan}
\date{April 15, 2004} 
\begin{abstract}
We have recently shown that dense quark matter
possesses a color ferromagnetic phase
in which a stable color magnetic field arises spontaneously.
This ferromagnetic state has been known to be Savvidy vacuum
in the vacuum sector.
Although the Savvidy vacuum
is unstable, 
the state is stabilized
in the quark matter.
The stabilization is achieved by the formation of quantum Hall states of gluons,
that is, by the condensation of the gluon's color charges
transmitted from the quark matter. 
The phase is realized between the hadronic phase and the color 
superconducting phase.
After a review of quantum Hall states of electrons in semiconductors,
we discuss the properties of 
quantum Hall states of gluons 
in quark matter in detail.
Especially, 
we evaluate the energy of the states as a function of
the coupling constant.
We also 
analyze solutions of vortex excitations
in the states
and evaluate their energies.
We find that the states become
unstable as the gauge coupling constant becomes
large, or the chemical potential
of the quarks becomes small, as expected. 
On the other hand, with the increase of the chemical potential,
the color superconducting state arises
instead of the ferromagnetic state.
We also show that the quark matter produced by
heavy ion collisions 
generates observable 
strong magnetic field $\sim 10^{15}$ Gauss 
when it enters the ferromagnetic phase.
\end{abstract}
\hspace*{0.3cm}

\maketitle
\section{introduction}
Quark matter is known or expected to have several phases,
hadronic phase, quark gluon plasma phase and color superconducting
phase\cite{color}. When the density of the quarks is small, the hadronic phase
arises in low temperature owing to very strong gluonic interactions. 
Thus, the quarks in the phase are confined\cite{monopole,abelian} in
the hadrons. On the other hand when the density of the quarks
is sufficiently large, the color superconducting phase is expected to arise.
In such a case, gluonic interactions are small, so that attractive
forces operate perturbatively between anti-triplet pair of the quarks.    
Thus, the condensation of the pairs arises to make the superconducting
phase realized. When the temperature is sufficiently high,
quark matter in both phases melts and forms the quark gluon plasma.
Among the phases only the hadronic phase is observed.
Although the color superconducting phase is very intriguing,
present experiments could not produce the phase because
large chemical potentials of the quark number such as $1$ GeV
is needed for the production of the phase.

We have recently discussed\cite{ferro} a possibility
of the stable color ferromagnetic states in dense quark matter.
The ferromagnetic state is caused by 
the condensation of the color magnetic field, not by
the  alignment of the quark's magnetic moments.
The states are realized 
between the hadronic state and the color superconducting state
when the chemical potential is varied. Thus, the phase 
could be observed in the present experiments.
The ferromagnetic states possess a spontaneously generated 
color magnetic field in maximal Abelian sub-algebra 
and also involve a quantum Hall state of off-diagonal gluons.
The gluons have been known to have unstable modes\cite{nielsen}
in the color magnetic field $B$. They occupy the lowest Landau level
with their spins pointed to the magnetic field and with their energies
being imaginary. The existence of these unstable modes 
implies that the naive ferromagnetic state ( Savvidy vacuum \cite{savidy}) is
unstable. We have recently shown\cite{ferro} that
the formation of the gluon's quantum Hall state ( QHS ) 
caused by the condensation of the unstable modes stabilizes 
the ferromagnetic state.

In this paper we review  
QHS\cite{qh} of electrons in semiconductor and Chern-Simons 
gauge theory\cite{iwa} for describing the QHS in the next section. 
This is because the phenomena and its theory are not popular
with hadron physicists. In the section \ref{qHs},
applying the 
theory to the unstable gluons, we discuss the properties of
QHS of the gluons; incompressibility, Laughlin quasiparticle, etc.
Especially, we numerically show that the energy of the Laughlin quasiparticle becomes
smaller as the gauge coupling constant becomes larger.
Since it may vanish at the infinite coupling constant,
a bound state of the quasiparticle and anti-quasiparticle     
would be a state with zero energy even at finite coupling constant.
This implies that the QHS of the gluons becomes unstable against
the creation of the bound states at the coupling constant. 
These excitations 
destroy  Laughlin state of the gluons.
As a result, 
the QHS decays and the ferromagnetic state also decays.
Instead, quark confining state would appear at such large coupling constant.
In other word, at small chemical potential
the hadronic phase arises instead of the ferromagnetic phase.
In the section \ref{cf} we show that the color superconducting state
is more favored than the ferromagnetic state when the number density 
of the quarks is sufficiently large. In the section \ref{obs} 
we consider phenomenological implications of the color
ferromagnetic states of the quark matter. Finally in the section \ref{dis}
we discuss that the existence of the ferromagnetic phase 
is a very natural consequence in the gluon and quark dynamics.

\section{quantum Hall state of electrons}
\subsection{Integer quantum Hall state}
QHS of electrons was discovered\cite{qhs} in 1980 by von Klitzing.
%
%
He has observed a plateau of Hall conductivities $\sigma_{xy}$ 
as a function of magnetic field imposed on the
two-dimensional quantum well fabricated in a semiconductor.
It implies that $\sigma_{xy}$ is quantized with the unit of 
the fundamental constant $e^2/2\pi\hbar$ at the plateau. 
It is called quantum Hall effect.
The observation indicated the existence of a specific state
of two-dimensional electrons in the well 
under the strong magnetic field, $B$,
typically $10^{5}$ Gauss. 
In these experiments,  
electrons are trapped in two-dimensional quantum well with its
width $\sim 10$ nm, so that their motions are restricted in two-dimensional space.
In order to move in a direction perpendicular to the space,
electrons need to gain energies $\sim 100$ eV. Thus, 
in experiments with low temperature $\sim 1$K, 
electrons move only in the two-dimensional space.

The two-dimensional electrons in the magnetic field 
make cyclotron motions with their radius $\ell_B=1/\sqrt{eB}$ and their
states are
specified by Landau levels. Each of them 
has a large number of degenerate states; the degeneracy   
per unit area is given by $eB/2\pi$. 
The original QHS was called integer quantum Hall state since
the state is observed at filling factor being integer;
the filling factor is defined as $\rho_{\rm e}/(eB/2\pi)$ ( $\rho_{\rm e}$ is two-dimensional number
density of electrons, typically $10^{11}/\mbox{cm}^2$. ) 
Thus, the filling factor 
means a fraction of electron occupation in a Landau level.
For example, the filling factor $\nu=1/3$ implies that electrons 
occupy a third of the lowest Landau level. 
The integer filling factor
implies that some of Landau levels are completely occupied.

Integer quantum Hall effect can be understood as a localization property 
of each two-dimensional electrons; some of them are
localized due to impurities and
the others are not localized in the magnetic field
in spite of the impurities. 
In general, all of two-dimensional electrons must be localized
around impurities. This is well known as Anderson localization.
Then, the system is an insulator because 
there are no carriers of electric currents;
localized electrons do not carry the currents.
But the localization theorem does not hold
when the magnetic field is present. 
We note that the effect of the impurities
lifts the degeneracy of the states in a Landau level.
Thus, the density of states 
has a finite width. Under the circumstance, 
almost of all the states 
are still localized. But,
electrons occupying the states around the center of the
Landau level
are extended all over the system so that they can carry electric currents.
This property of the localization in two dimensional electrons 
yields plateaus in the Hall conductivity around $\nu=$ integers.
In this way, integer quantum Hall effects are caused by the interplay of the
impurities and each electron, and
many-body correlations among electrons are not important.

\subsection{Fractional quantum Hall state}

Fractional quantum Hall effects
were discovered\cite{tsui} in 1982 by Tsui at the filling factor
being fractional numbers, e.g.~$1/3,\,2/3 $.
( He observed quantum Hall plateaus at such fractional filling factors. )
Electrons occupy a fraction in the lowest Landau level.
The fractional QHSs are understood to be 
caused by many-body effects of electrons,
just like superconductivity. Impurities do not play 
important roles in these QHSs.
Laughlin\cite{laughlin} proposed a wavefunction for this QHS,
at the filling factor being $1/n$,
\begin{equation}
\Psi=\prod_{i,j}(z_i-z_j)^n \exp(-eB\sum_i|z_i|^2/4)
\end{equation}
with $z=x+iy$ denoting complex coordinate of electrons
with charge $-e$,
where we have used a symmetric gauge potential $\vec{A}^B=(yB/2,-xB/2,0)$
for the magnetic field.
Here, $n$ is an odd integer for the Fermi statistics of electrons.
Numerical simulations
show that the ground states of the electrons at the fractional filling
factors are well described by the Laughlin
wavefunctions even if the repulsive Coulomb interaction 
is replaced by a delta function; the precise form of the 
interaction between electrons is not important for the realization
of QHS.

In general, a system of electrons partially occupying the lowest Landau
level is compressible, namely, the system has no
gap; excitation levels are distributed continuously above
the ground state. However, the QHS has a gap
just like the BCS state. 
That is, the QHS is characterized as a state with gap.
We note that free electron gas has no gap, so that even with
the Coulomb interaction taken into account the gas does not gain the
gap in general. But
the BCS states are gapped states
formed from Fermi gas with small attractive force
among electrons around Fermi surface. Similarly, the QHSs 
are gapped states formed from the gas of two-dimensional electrons interacting 
repulsively with each other under the 
strong magnetic field. These gapped states arise at the fractional
filling factors.

The gap of the QHS is given by the energy of the bound state of the
Laughlin's quasiparticles. The quasiparticles were originally discussed by Laughlin
based on the Laughlin wave functions\cite{qh}. They were argued to possess
fractional electric charges $\pm e/n$ in the QHS with the filling factor, $\nu=1/n$. Namely,
If we add an electron to the QHS,
there appear $n$ Laughlin's quasiparticles each of which possesses 
a fractional charge of $-e/n$. On the other hand, if we extract an electron from
the state, there appear $n$ Laughlin's anti-quasiparticles, each of which 
possesses a fractional charge of $e/n$. These quasiparticles,
not electrons, play the role of carrying electric currents in the fractional QHS.
Therefore, the Hall conductivity is given by $e/2\pi\times e/n$. 
The plateaus at the fractional filling factor are understood\cite{qh} as
the localization properties of these quasiparticles.
The excitations on the QHS are
given by bound states of the quasiparticle and anti-quasiparticle;
they attract each other because of the Coulomb interaction.
In this way the gap of the QHS is given by the energy of this bound state of the
Laughlin's quasiparticles.
 
\subsection{Chern-Simons gauge theory of quantum Hall state}

We have BCS theory for understanding superconducting states.
But we do not have similar theory for the QHS. Namely, there is no
effective fermionic theory of electrons explaining the properties of the QHS, e.g.
Laughlin wavefunctions.
We only know that numerical simulations using Hamiltonian of electrons interacting with
each other through the Coulomb potential, confirm the validity of the wavefunctions
and the gap in the state.
On the other hand, there is a bosonic theory in the case of the superconductivity, 
which is well known as Landau-Ginzburg theory. Similarly,
there is a Chern-Simons gauge theory of bosonized electrons
for understanding the QHS. In this section, we wish to explain it.

It is well known that BCS states are described by 
Landau-Ginzburg effective Lagrangian,

\begin{equation}
\label{BCS}
L_{\rm BCS}=|(i\partial_{\mu}+2eA_{\mu})\phi|^2+m^2|\phi|^2-\lambda |\phi|^4-\frac{1}{4}F_{\mu\nu}F^{\mu\nu}
\end{equation}
where $\phi$ and $A_{\mu}$ denote Cooper pair of electrons
and electromagnetic fields respectively. The ground state is a condensed state of the Cooper pairs,
namely, $\langle\phi\rangle=\sqrt{m^2/2\lambda}$. 
Since the gauge symmetry, $\phi\to\phi
e^{i\Lambda(x)}$ and $A_{\mu}\to A_{\mu}+\partial_\mu\Lambda/2e$, is 
spontaneously broken, vortex excitations with the magnetic flux
$2\pi/2e$ arise. These are magnetic vortices penetrating superconductors.
When we switch off the gauge interaction, we have an effective model 
of superfluids.

Similarly, there is a bosonic theory for the QHS of electrons.
It is a theory of composite electrons with
Chern-Simons gauge field\cite{zhang,iwa},

\begin{equation}
L_{\rm QHS}=\phi_{\rm e}^{\dagger}(i\partial_0-a_0)\phi_{\rm e}+{\rm c.c.}
-\frac{1}{2m_{\rm e}}|(i\partial_i+eA_i^B-a_i)\phi_{\rm e}|^2-V_{\rm Coulomb}
+\frac{1}{4\alpha}\epsilon^{\mu\nu\lambda}a_{\mu}\partial_{\nu}a_{\nu}
\end{equation} 
where $m_{\rm e}$ denotes mass of electrons and $V_{\rm Coulomb}
=\int d^2xd^2y(e|\phi_{\rm e}(x)|^2-\bar{\rho})\frac{1}{2|x-y|}(e|\phi_{\rm e}(y)|^2-\bar{\rho})$ 
is the Coulomb interaction between electrons with background positive charges $\bar{\rho}$. The term
$A_{\mu}^B$ denotes the external magnetic field imposed for
the realization of QHS and has no kinetic term. The factor of $\alpha$
should be taken as $\pi\times $odd integer for the boson field $\phi_{\rm e}$
to describe fermionic electrons.  

The boson field $\phi_{\rm e}$ represents composite electrons;
boson $\phi_{\rm e}$ attached with flux of $a_i$.
That is, fermions in two-dimensional space can be described by
bosons attached with a fictitious flux $2\alpha$ of Chern-Simons
gauge field $a_i$. Owing to this flux, the exchange of the bosonic
particles induces a phase $e^{i\alpha}$ in their wavefunction. Thus,
with the choice of $\alpha=\pi\times $odd integer,
the wavefunctions represent particles with Fermi statistics. 
One can show that $L_{\rm QHS}$ serves to describe the composite electrons.

Using Hamiltonian derived from the Lagrangian,
we can obtain Schr\"{o}dinger equation for electrons with Fermi
statistics. In that sense, the Lagrangian correctly describes
two-dimensional system of electrons in the magnetic field.
We should note that if Chern-Simons gauge fields are absent in the
above Lagrangian and $\phi_{\rm e}$ obeys Fermi statistics, the Lagrangian
describes ordinary electron system.

We should comment that the field $\phi_{\rm e}$ describes
an electron with a particular spin component parallel to the magnetic
field. In order to describe the electron
with the other spin component,
we need an additional boson field as well as Chern-Simon gauge fields.

>From equations of motions derived from $L_{\rm QHS}$, 
we can see that QHSs are ground-state solutions such
as $\langle\phi_{\rm e}\rangle\neq 0$ similar to the case of BCS states.
These solutions can be obtained only when the relation, $eA_i^B=a_i$, holds.
Namely, the magnetic field $eA_i^B$ is canceled 
by the Chern-Simons gauge
field $a_i$. The Chern-Simons flux $\epsilon_{ij}\partial_ia_j$ can be represented by the density $\rho_{\rm e}$ of electrons
$\phi_{\rm e}^{\dagger}\phi_{\rm e}$ such as
$\phi_{\rm e}^{\dagger}\phi_{\rm e}=-\epsilon_{ij}\partial_ia_j/2\alpha$,
an equation derived by taking variational derivative of $L_{\rm QHS}$ in
$a_0$. Hence, the solutions can be found only when the filling factor
$\nu=2\pi\rho_{\rm e}/eB$ is given by $\pi/\alpha$. In this way we can understand that QHSs
are condensed states of bosonized electrons $\phi_{\rm e}$ 
and are realized only at $\nu=1/3,1/5$, etc.~for $\alpha=3\pi,5\pi$, etc. 

In order to see\cite{zhang} that the states are really QHSs with appropriate Hall
conductivities, $\sigma_{xy}$,
we derive $\sigma_{xy}$ in the following.
We introduce a gauge potential $A_{\mu}$ of electric field
$\vec{E}=-\partial_0\vec{A}-\vec{\partial}A_0$ in $L_{\rm QHS}$;
$\phi_{\rm e}^{\dagger}(i\partial_0-a_0+e A_0)\phi_{\rm e}+ \cdots$ .
Shifting the integration variable of $a_{\mu}$ in the functional integral 
$Z(A_{\mu})=\int D\phi_{\rm e} Da_{\mu}\exp(i\int d^2xdt L_{\rm QHS})$ such as
$a_{\mu}\to a_{\mu}+eA_{\mu}$, we calculate electric current
$j_x=-i\partial_{A_x}\log Z$, which is given by $e^2E_y/2\alpha$
in the state of $\langle\phi_{\rm e}\rangle=\sqrt{\rho_{\rm e}}$ and 
$\langle-\partial_0a_y-\partial_ya_0\rangle=0$. 
Thus, the Hall conductivity
is correctly given by $\nu e^2/2\pi$.
In this way we find that the condensed state
of the bosonized electrons is the QHS.  
( One ( A.I. ) of the authors previously showed\cite{iwa2} that Laughlin wavefunctions
can be derived from the condensed states of the field $\phi_{\rm e}$. )

Laughlin's quasiparticles are excited states of the QHS.
In the picture of the bosonized electrons,
they are presented by vortex excitations on the state of
$\langle\phi_{\rm e}\rangle=\sqrt{\rho_{\rm e}}\neq 0$, 
where U(1) gauge symmetry ( $\phi_{\rm e}\to \phi_{\rm e} e^{i\Theta}$
and $a_{\mu}\to a_{\mu}-\partial_{\mu}\Theta$ )
is spontaneously broken and hence there are topological excitations
associated with the symmetry. 
Actually, we can find a vortex soliton such that
$\phi_{\rm e}(x)=f(r)\exp(i n \theta)$ with boundary conditions, $f(r)\to
\sqrt{\rho_{\rm e}}$ and $a_i\to -\partial_i (n\theta)+eA_i$
as $r\to \infty$ and $f(r=0)=0$, where $\theta$ is an azimuthal angle and $n$ is an integer.
This vortex is similar to 
the magnetic vortex in the superconductor, but 
the former has a quantized electron number while the latter has the 
quantized magnetic flux.
This is because the flux quantization $-\int d^2 x
\epsilon_{ij}\partial_ia_j=2\pi n $
implies the electron number quantization $N_{\rm e}=-\int d^2 x
\epsilon_{ij}\partial_ia_j/2\alpha=\pi n/\alpha$ of the vortex solutions.
Thus, Laughlin's quasiparticles have a fractional electric charge
such as $e/3$. 
This fractionality of the electric charges has been observed\cite{fruct}.
In this way the theory of the composite electrons
can describe
the QHSs as condensed states of bosons in the mean field approximation
just like Landau-Ginzburg
theory of superconducting states. ( This similarity  
can naturally lead to a prediction of 
the presence of Josephson effects in bilayer quantum
Hall systems\cite{jo}.)

\section{quantum Hall state of gluons} \label{qHs}
\subsection{Unstable gluons in color magnetic field}

Up to now, we have given a brief review of the theory of QHSs in the 
two-dimensional electron system.
We, now, apply the idea to analyze a QHS of gluons, which appears
in dense quark matter. We discuss SU(2) gauge theory for simplicity.

It has been known\cite{savidy} that in the gauge theory one-loop effective potential for color magnetic field
has non-trivial minimum; $V(g{\cal B})= \frac{11}{48\pi^2}g^2{\cal B}^2\left(
\log(g{\cal B}/\Lambda^2)-\frac{1}{2}\right)-\frac{i}{8\pi}g^2{\cal B}^2$,
with an appropriate renormalization of the gauge coupling $g$.
Here, we have not included contributions from quarks. Even if their
contributions are included, only change in $V(g{\cal B})$ is a numerical factor
in the coefficient of the first term. 
( Beyond the one-loop approximation, the presence of the nontrivial 
minimum in $g{\cal B}$ has been proved\cite{nn} in general under the 
assumption that the running coupling constant $g(g{\cal B})$ becomes
infinity at a finite $g{\cal B}$. )
This apparently seems to imply spontaneous generation of the color magnetic field, namely,
the realization of a ferromagnetic state. But it is not so simple since 
the imaginary part in $V(g{\cal B})$ is present when $g{\cal B}\neq 0$.
It means that the state with the magnetic field  is unstable
as well as the
perturbative vacuum state
with $g{\cal B}=0$.
Actually, the unstable modes of gluons are present\cite{nn} in the state with the magnetic field
and are
expected to make some stable condensed states.
What kind of the stable state is formed of the unstable gluons?
We have shown\cite{ferro} that the state is a QHS of the gluons with the color magnetic field.
In order to explain it, we rewrite the gluon's
Lagrangian
with the use of the variables, ``electromagnetic field" 
$A_{\mu}=A_{\mu}^3,\,\,\mbox{and} \,\, \mbox{``charged vector field"}\,
\Phi_{\mu}=(A_{\mu}^1+iA_{\mu}^2)/\sqrt{2}$ 
where indices $1\sim 3$ denote color components,

\begin{eqnarray}
\label{L}
L=-\frac{1}{4}\vec{F}_{\mu
  \nu}^2&=&-\frac{1}{4}(\partial_{\mu}A_{\nu}-\partial_{\nu}A_{\mu})^2-
\frac{1}{2}|D_{\mu}\Phi_{\nu}-D_{\nu}\Phi_{\mu}|^2 \nonumber \\
& & \ \ +ig(\partial^{\mu}A^{\nu}-\partial^{\nu}A^{\mu})\Phi_{\mu}^{\dagger}
\Phi_{\nu}+\frac{g^2}{4}(\Phi_{\mu}\Phi_{\nu}^{\dagger}-
\Phi_{\nu}\Phi_{\mu}^{\dagger})^2
\end{eqnarray}
with $D_{\mu}=\partial_{\mu}+igA_{\mu}$.
We have used a gauge condition, $D_{\mu}\Phi^{\mu}=0$. 
Using the Lagrangian we can derive that the energy $E$ of
the charged vector field $\Phi_{\mu}\propto e^{iEt}$ 
in the magnetic field, $A_{\mu}=A_{\mu}^B$, is given by
$E^2=k_3^2+2g{\cal B}(n+1/2)\pm 2g{\cal B}$  
with a gauge choice, $A_j^B=(0,x_1 {\cal B},0)$ and $(\partial_{\mu}+igA_{\mu}^B)\Phi^{\mu}=0$, 
where we have taken the spatial direction of $\vec{\cal B}$ being along $x_3$ axis.
$\pm 2g{\cal B}$ expresses the contribution from spin components
of $\Phi_{\mu}$. The integer $n\geq 0$ 
and $k_3$ represent Landau level and 
momentum in the direction parallel to the magnetic field, respectively.

Obviously, the modes with $E^2(n=0)<0$ are unstable modes
occupying the lowest Landau level and with spin parallel to $\vec{\cal B}$. 
Among them, the modes with $k_3=0$ are the most unstable ones, which
means that they have
the largest negative value of $E^2(k_3=0)$.
Thus, they are expected to play the main role of forming a stable state. 
Here we should
remember a simple model of a scalar field with 
 a double-well
potential, $-m^2|\phi|^2+\lambda |\phi|^4/2$. 
The state $\langle\phi\rangle=0$ is unstable and unstable modes
$\phi(\vec{k})$ with their energies $E^2=\vec{k}^2-m^2<0$ arise on the
state. 
In the case,
the most unstable uniform mode, $\phi(\vec{k}=0)$, among $\phi(\vec{k})$,
condenses to form a stable state $\langle\phi\rangle=\sqrt{m^2/\lambda}$.
Therefore, the most unstable modes with $k_3=0$ 
are relevant to the formation of the true ground state also in the gauge theory. 
Since they have no $x_3$ dependence, they are two-dimensional objects occupying the lowest Landau
level. The situation is quite analogous to the case in 
the two-dimensional electrons forming QHSs just as mentioned above. 
The only difference is that 
in the gauge theory gluons are bosons, while electrons are fermions.

In order to find the stable state in the gauge theory,
we extract only the most unstable modes from the Lagrangian, eq.~(\ref{L}),
ignoring the other modes coupled with them and 
obtain a two-dimensional Lagrangian,

\begin{equation}
L_{\rm unstable}=|(i\partial_{\nu}-gA_{\nu}^B)\phi_{\rm u}|^2+2g{\cal B}|\phi_{\rm u}|^2-\frac{\lambda}{2} |\phi_{\rm u}|^4, 
\end{equation}
with $\lambda=g^2/\ell$, where the field
$\phi_{\rm u}=(\Phi_1-i\Phi_2)\sqrt{\ell/2}$
denotes the unstable modes in the lowest Landau level.
(We used the condition that these modes occupy the lowest Landau 
level: $(D_1+iD_2)\phi_{\rm u}=0$.)
$\ell$ is the coherent
length of the magnetic field. 
Here we are thinking the quark matter with its length scale $\ell$.
Then,  a condition of $\ell\gg \ell_B=1/\sqrt{g\cal B}$ must be satisfied for the consistency.
Without the condition, the states of quarks and gluons can not be specified by Landau levels. 
We note that the field $\phi_{\rm u}$ has a color charge associated with
$\tau_3$ of SU(2) algebra. This color charge is only a conserved quantity
when the color magnetic field $\propto
\tau_3$ is generated spontaneously in the SU(2) gauge theory.

This Lagrangian is quite similar
to the Lagrangian in eq.~(\ref{BCS}) of the superconductivity.
It apparently seems that the ground state is simply given by
$\langle\phi_{\rm u}\rangle= \sqrt{2g{\cal B}/\lambda}$, the condensed state of the field $\phi_{\rm u}$. 
But it is impossible because the term of $A_{\mu}^B$ is present in the
kinetic term. If this term vanishes, the term of the negative mass
also vanishes so that the solution $\langle\phi_{\rm u}\rangle\neq 0$ does not
exist. Physically, the Lagrangian $L_{\rm unstable}$ describes such a system that 
the particles of $\phi_{\rm u}$ move in the magnetic field and interact with
each other through a repulsive potential of a delta function.
There is a numerical simulation\cite{nakajima} that the nonrelativistic particles 
with such a repulsive potential
can form a Laughlin state even if they are bosons.
Thus, the gluons represented by $\phi_{\rm u}$ may form a quantum Hall state.

\subsection{Quantum Hall state of unstable gluons}

In order to see the QHS of the field, $\phi_{\rm u}$, explicitly, 
we introduce Chern-Simons gauge field to make composite gluons;
bosons attached with the Chern-Simons flux. Then, a relevant
Lagrangian is given by 

\begin{equation}
\label{la}
L_a=|(i\partial_{\nu}-gA_{\nu}+a_{\nu})\phi_a|^2+2g{\cal B}|\phi_a|^2-\frac{\lambda}{2}|\phi_a|^4+
\frac{\epsilon^{\mu\nu\lambda}}{4\alpha}a_{\mu}\partial_{\nu}a_{\lambda},
\end{equation}
where the statistical factor $\alpha$
should be taken as $\alpha=2\pi\times $integer to maintain the
equivalence of the system described by $L_a$ to that of
$L_{\rm unstable}$. This new field $\phi_a$ represents the composite gluons
attached with the Chern-Simons flux $a_i$.

The equivalence between $L_{\rm unstable}$ and $L_a$
has been shown \cite{seme} in the operator formalism although the
equivalence had been known in the path integral formalism using the world
lines of the $\phi_a$ particles. ( In the formalism the last term in eq.~(\ref{la}) produces
a phase, $e^{i\alpha/\pi}$, in wavefunctions when trajectories of two particles are
interchanged. )
This Lagrangian corresponds to $L_{\rm QHS}$ of composite electrons. 
Obviously, there is the U(1) gauge symmetry such that
$\phi_a \to \phi_a e^{i\Lambda}$ and $a_{\mu}\to
a_{\mu}+\partial_{\mu}\Lambda$; a nonvanishing term, 
$\frac{\epsilon^{\mu\nu\lambda}}{4\alpha}\partial_{\mu}\Lambda\partial_{\nu}a_{\lambda}$ 
in $L_a$ under the gauge transformation vanishes in the action integral
$\int d^3xL_a$ with appropriate boundary conditions.

In deriving equations of motion,
we need to impose a condition of the modes $\phi_a$
occupying the lowest Landau level\cite{iwa3}.
The condition is given by $(D^a_1+iD^a_2)\phi_a=0$ with $iD^a_i=i\partial_i-gA_i+a_i$.
Thus, adding a term
$C(D^a_1+iD^a_2)\phi_a$ to $L_a$ with a Lagrange multiplier $C$,
we derive equations of motion by
taking functional derivatives in $\phi_a$, $a_{\mu}$ and $C$,

\begin{eqnarray}
\label{eq1}
&&\phi_a^{\dagger}\,i\partial_0\,\phi_a+{\rm  c.c.}+2a_0\,|\phi_a|^2=-\frac{1}{2\alpha}\epsilon_{ij}\,\partial_i\,a_j
\\
\label{eq2}
&&-\epsilon_{ij}\partial_j|\phi_a|^2+\delta_{i1}i(-C\phi_a+C^{\dagger}\phi_a^{\dagger})
+\delta_{i2}(C\phi_a+C^{\dagger}\phi_a^{\dagger})
=\frac{1}{2\alpha}\epsilon_{ij}(\partial_0\,a_j-\partial_i\,a_0) 
\\
\label{eq3}
&&(i\,\partial_0+a_0)^2\,\phi_a+(g{\cal B}-\epsilon_{ij}\partial_ia_j)\phi_a
-(D^a_1-iD^a_2)C^{\dagger}=\lambda
|\phi_a|^2\phi_a \\
&&(D^a_1+iD^a_2)\phi_a=0
\end{eqnarray}
where we have used a formula\cite{iwa3} of $\int d^2x|D^a_i\phi_a|^2
=\int d^2x
(|D^a_1+iD^a_2)\phi_a|^2+(g{\cal B}+\epsilon_{ij}\partial_ia_j)|\phi_a|^2)$;
surface terms are omitted in this formula.

We find that the solution of the uniform ground state is
given such that $C=0$, $a_i=gA_i^B$ and, $a_0$ and $\phi_a$
are solutions of the equations, 

\begin{equation}
2a_0|\phi_a|^2=\frac{g{\cal B}}{2\alpha} \quad \mbox{and}\quad
a_0^2+2g{\cal B}=\lambda |\phi_a|^2.
\end{equation}
That is, the QHS represented by the condensed state,
$\langle\phi_a\rangle=v\neq 0$,
arises only when the magnetic field is canceled by
the Chern-Simons field,
$-\epsilon_{ij}\partial_ia_j=g{\cal B}=2\alpha\rho_{\rm c}$;
$\rho_{\rm c}$ is given by the left hand side of eq.~(\ref{eq1}), i.e. 
$\rho_{\rm c}=2a_0v^2$. This $\rho_{\rm c}$ represents 
color charge density possessed by the gluons $\phi_a$.
The composite gluons condense to form the QHS only when
$\nu=2\pi\rho_{\rm c}/g{\cal B}$ is equal to $\pi/\alpha$.  
This is quite similar to the case of the ordinary QHS
mentioned above.
It is easy to show that this state possesses appropriate Hall
conductivity $\sigma_{xy}=(\pi/\alpha)g^2/2\pi$.
Therefore, we understand that the condensed state $\langle\phi_a\rangle\neq 0$
is a QHS of gluons.
It apparently seems that there are infinitely many QHSs with the
filling factor $\nu=\pi/\alpha$ because $\alpha$ can take 
infinitely many values such as 
$n\times \pi$ with positive even integer $n$. In the QHSs of
electrons, the states with small filling factors have
low densities of electrons. In such a case electrons forming 
Wigner crystal is energetically more stable than electrons forming the QHSs.
Actually, such QHSs with small filling factors, e.g.~$1/9$, have not
been observed. We expect that similarly in the gauge theory, 
Wigner crystal of gluons would be realized when 
the filling factor is much small $\nu \ll 1$.
The analysis is now in progress. 

We should mention that the QHS of gluons is realized in a sector with
nonzero color charge, not in the vacuum sector; the condensed state of $\phi_a$ possesses a
color charge. Such a state can arise in dense quark matter where
the color charge of quarks is transmitted to the condensate.
This fact leads to the minimum number density of quarks
for realizing a QHS, for example, QHS with $\nu=1/2$
where the color charge density of the condensate given by $g{\cal B}/4\pi\ell$ 
must be supplied by the quarks. Since the color charge of the quarks
is a half of the gluon's, the necessary number density $\rho_{\rm q}$ of the quarks 
for producing the QHS of the gluons is given by 

\begin{equation}
\rho_{\rm q}=\rho^{(+)}+\rho^{(-)}=2\rho^{(+)}=2\,n_{\rm f}\, n_{\rm c}
\int\frac{ d^3k}{(2\pi)^3}=\frac{n_{\rm f}\,n_{\rm c}k_{\rm f}^3}{3\pi^2}
=\frac{4k_{\rm f}^3}{3\pi^2}=\frac{g{\cal B}}{4\pi \ell}
\end{equation}
with number density $\rho^{(\pm)}$ of positive ( negative ) colored quarks,
where $n_{\rm f}=2$ and $n_{\rm c}=2$ are the number of flavors and colors
respectively.
$k_{\rm f}$ denotes the Fermi momentum given by $\sqrt{\mu^2-m_{\rm q}^2}$
 with the quark mass $m_{\rm q}$ and the chemical
potential $\mu$ of the quark matter at zero temperature. Therefore, it turns out that
the minimum chemical potential $\mu$ for realizing the QHS is given by
$\sqrt{(3\pi g{\cal B}/16\,\ell)^{2/3}+m_{\rm q}^2}$. We note that this value of $\mu$
is necessary, not sufficient for the realization of the state.
In this way, the presence of the dense quark
matter is necessary for producing the QHS of gluons. 
On the other hand, when we are concerned with the vacuum sector, such a QHS cannot arise
so that the ferromagnetic state is unstable. Probably, the large
fluctuation of
the unstable modes may form a confining vacuum
called a spaghetti vacuum \cite{spa}.

In order to calculate the ground state energy  
we derive Hamiltonian,

\begin{equation}
\label{H}
H=\int d^2x \left(
 a_0^2|\phi_a|^2+(g{\cal B}+\epsilon_{ij}\partial_ia_j)|\phi_a|^2-2g{\cal B}|\phi_a|^2
+\frac{\lambda}{2}|\phi_a|^4 \right)
\end{equation}
Thus, the energy density $E_2(v)$ of the QHS is given by $E_2(v)=a_0^2v^2-2g{\cal B}v^2+\frac{\lambda}{2}v^4 $.        
We should note that $E_2(v)$ represents the  energy density in two-dimensional space
and that the three-dimensional one is given by $E_2(v)/\ell$.

The behavior of the ground state solution with respect to 
the coupling $\lambda =g^2/\ell$ and the filling factor 
$\nu=\pi/\alpha$ is given by 

\begin{eqnarray}
&v&\to \left(\,\frac{g\cal B}{4\alpha\sqrt{\lambda}}\,\right)^{1/3}\, , \,\,
\quad a_0\to \left(\,\frac{g\cal B\lambda}{4\alpha}\,\right)^{1/3}\,
\,\, 
\quad \mbox{for}\quad \lambda\to\infty  \nonumber \\
&v&\to\sqrt{\frac{2g\cal B}{\lambda}}\quad , \quad\,\,\,a_0\to\frac{\lambda}{8\alpha}
\hspace{7em} \mbox{for}\quad \frac{\lambda}{\alpha} \ll \ell_{B}^{-1}
\quad \mbox{respectively.} 
\end{eqnarray} 
Thus, we find that the ground state energy density ( $=E_2(v)/\ell+{\rm Re}
V(g{\cal B})$ ) in three-dimensional space
becomes large such as
$E_3\sim 1.5\lambda^{1/3}(g{\cal B}/4\alpha)^{4/3}/\ell$ as the gauge coupling constant becomes large,
$\lambda\to \infty$ ( or as the length scale of the system in the direction 
of the magnetic field becomes
small, $\ell \to 0$ ).
The fact implies that the QHS becomes unstable as the coupling becomes large.
This is because the energy of the ferromagnetic state ( ${\cal B}\neq 0$ ) involving the
QHS becomes larger than the energy of the perturbative ground state
with ${\cal B}=0$; we have normalized the energy such that the energy of the perturbative
ground state vanishes at $g{\cal B}=0$.
This is consistent with naive expectation that at sufficiently large
$g^2$, 
the hadronic state ( ${\cal B}=0$ ) is realized instead 
of the ferromagnetic state: the hadronic or confining ground state 
is more stable than the
perturbative ground state for such a large coupling.
 Therefore, at large coupling constants the ferromagnetic state
becomes unstable and the hadronic state would be realized.

On the
other hand, $E_3\sim -0.5(2g{\cal B})^2(\lambda \ell)^{-1}$ as $\lambda\to 0$.
This implies that when the coupling constant is sufficiently small,
the ferromagnetic state is stable since it has much small energy.
It apparently seems to be unnatural because the perturbative ground
state may be realized at the small coupling. But we should mention
that the QHS of the gluons is realized only in dense quark matter,
not in the vacuum because 
for the realization of the QHS, the color charge associated with $\tau_3$
must be supplied from somewhere in the neutral system: 
The condensate of $\phi_a$ possesses the color charge,
which must be supplied from the quark matter.
Therefore, even at small coupling constants, the QHS can arise
as a stable state 
in the quark matter. ( In the vacuum the perturbative ground state is realized 
at such small coupling since 
there are no color charges. )

Analyzing small fluctuations $\delta\phi_a$, etc.~around the solution of the ground state, 
we can see that the energy of the fluctuations has a real positive gap given by
$\sqrt{4a_0^2+2\lambda v^2}$. The fluctuations represent extended collective motions, while there
are individual localized collective motions, namely Laughlin's
quasiparticles. They are vortex topological solitons in the
Chern-Simons gauge theory. We find from numerical analysis of such
solutions that the energies of the solitons
are positive. Therefore, no instability in the ferromagnetic state
( ${\cal B}\neq 0$ ) appears as a result of the formation of the QHS of the
gluons. In the next subsection we discuss the vortex solitons in the QHS of the gluons in detail.

\subsection{Vortex excitations in the quantum Hall state}

Such vortex solitons arise owing to the spontaneous breakdown of the U(1) gauge symmetry 
of the Lagrangian in eq.~(\ref{la}) describing spatially two-dimensional
gluons. The solutions can be obtained in the following.
First, we solve the lowest Landau level condition,
$(D^a_1+iD^a_2)\phi_a=0$. Then, we obtain $\phi_a=f(z)e^{a}$
where $f(z)$ is an arbitrary function of $z=x+i y$ and
$a$ is defined by putting $a_i=gA_i+\epsilon_{ij}\partial_ja$.
We assume that the solutions are spherically symmetric, namely,
$a(r)$ and $a_0(r)$ are functions only of the radial coordinate $r=\sqrt{x^2+y^2}$. 
Then, when we take $f(z)=vz^n=vr^ne^{in\theta}$, it represents
a solution of a vortex with vorticity being equal to $n$ where $n$ is a
positive integer. For simplicity, we consider only a solution with
$n=1$. Boundary conditions are such that $re^{a(r)}\to 1$ ( or $\phi_a\to v$ ) and
$a_0(r)\to a_0$ as $r\to
\infty$, and $re^{a(r)}\to 0$ as $r\to 0$ for avoiding a singularity at $r=0$.
This boundary condition at $r=\infty $ leads to a quantization of color charges
carried by the topological soliton. Namely, the soliton has
the flux $\int d^2x (-\epsilon_{ij}\partial_i a_j-g{\cal B})=\int d^2x \partial^2 a=2\pi$
where we have used the boundary condition. This means that
the color charge of the soliton, which is the integral of the color charge density given in 
the left hand side of eq.~(\ref{eq1}), is given by $\pi/\alpha$. In general it is given by
$n\times \pi/\alpha$ for the soliton with the vorticity of $n$. Therefore, we 
find that the color charge of the soliton is quantized.

Taking $C=\bar{z}^2b(r)$ with $b(r)$ being function only of $r$,
eqs.~(\ref{eq1})$\sim$(\ref{eq3}) are reduced to
\begin{eqnarray}
&&2a_0(r)v^2r^2e^{2a(r)}=\frac{g{\cal B}+\partial^2a(r)}{2\alpha} \\ 
&&a_0(r)^2+2g{\cal B}+\partial^2a(r)-\frac{1}{2v^2r^2e^{2a(r)}}
\partial^2\left(\frac{a_0(r)}{2\alpha}-v^2r^2e^{2a(r)}\right)=\lambda 
v^2r^2e^{2a(r)}.
\end{eqnarray} 
The second equation can be obtained
directly by inserting a solution $\phi_a=v ze^{a(x)}$ of the lowest
Landau level condition into the Hamiltonian, eq.~(\ref{H}), and by taking a
variational derivative in $a(x)$.  
We have solved the equations numerically, and obtained their configurations and energies
for various coupling parameter $\lambda=g^2/\ell$ and $\alpha$; Fig.~1 and Fig.~2.

We can see that the energy of the vortex soliton approaches zero
as $\lambda$ goes to infinity. We note that
the typical energy scale of the solution is given by $v^2(\lambda)$,
which goes to zero as $\lambda\to \infty$. Thus, the energy of the solution
goes to zero as $v^2(\lambda)\to 0$.
This indicates a possibility that the QHS becomes
unstable at sufficiently large coupling constant of $\lambda$ since the energy of the bound state
of a vortex ( $n=1$ ) and an anti-vortex ( $n=-1$ ) can
become negative at large coupling constants. They have
color charges opposite with each other and 
their binding energy
becomes larger than the intrinsic energies of the vortices at
large coupling constants. 
Then, such excitations of the bound states 
are produced unlimitedly and consequently, the QHS decays:
The whole space is occupied by such solitons and the condensate
melts because the condensate of the gluons vanishes at the center of the vortex; $\phi_a(r=0)=0$.  
This bound state
corresponds to the roton excitation in the QHS of electrons. 
Although these arguments are speculative,
the instability of the QHS discussed here is consistent with
the instability derived from the consideration of the QHS's energy
in $\lambda$.

\begin{figure}[t!]
\centering
\begin{minipage}{0.48\textwidth}
\includegraphics[width=\textwidth]{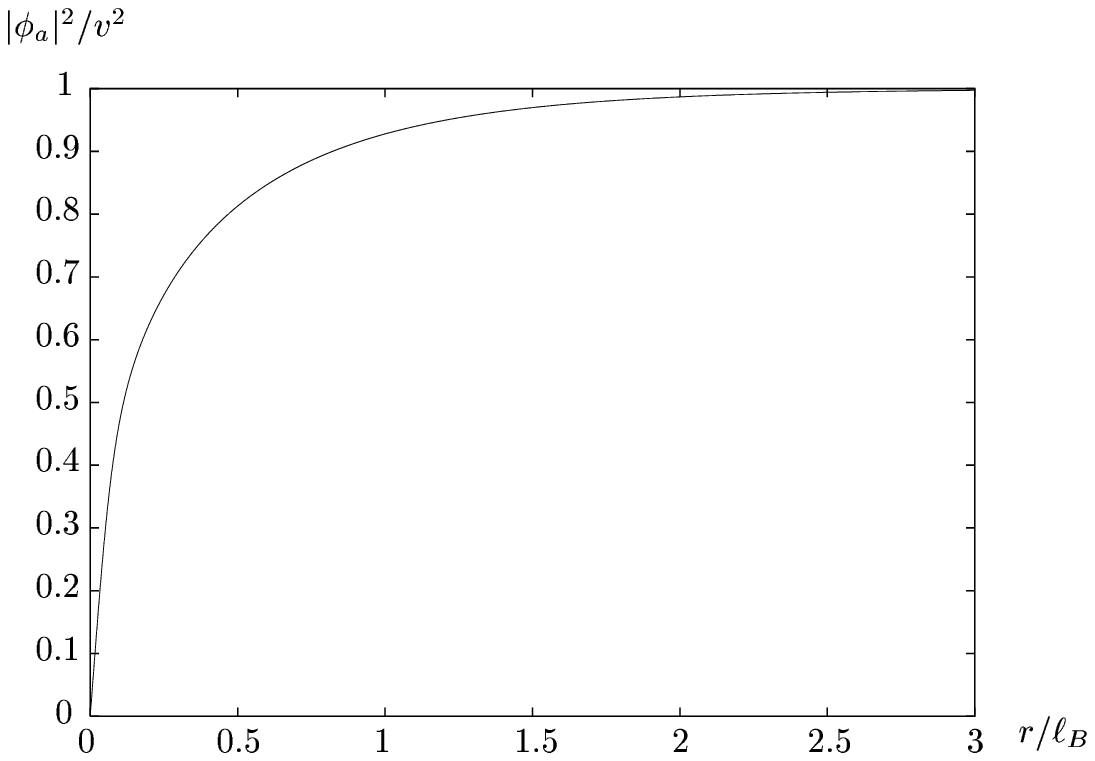}
\caption{Profile of the vortex}  
\label{fig:1}
\end{minipage}
\hfill
\begin{minipage}{0.45\textwidth}
 \centering
\includegraphics[width=\textwidth]{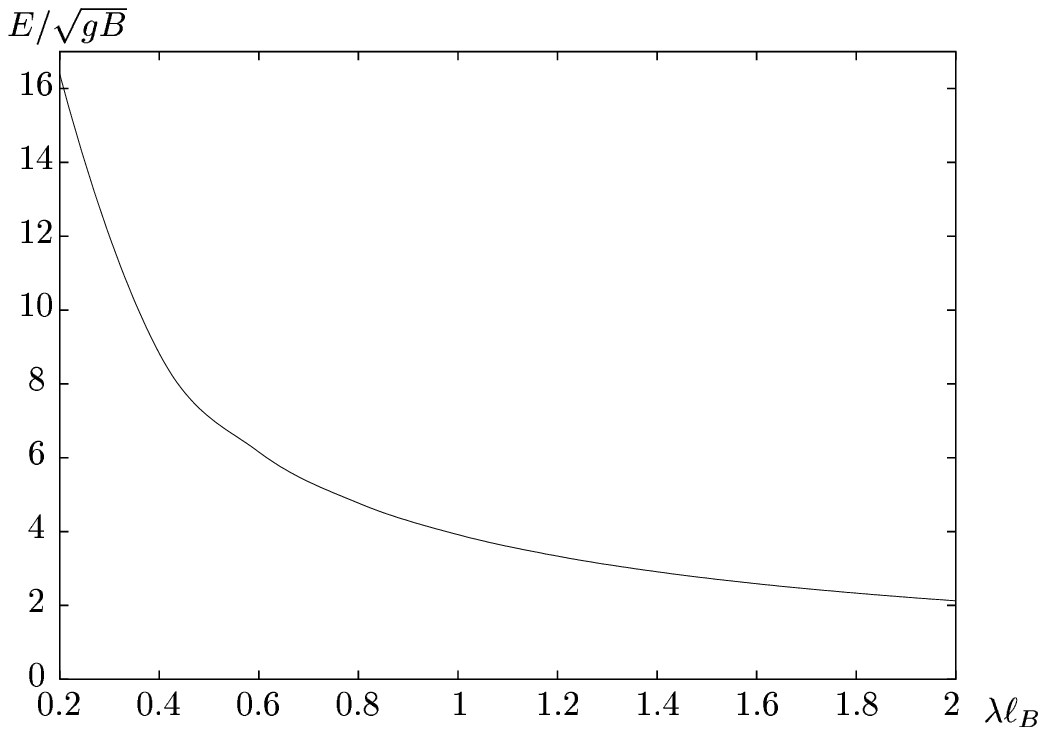}
\caption{Coupling dependence of the vortex excitation energy}
 \label{fig:2}
\end{minipage}
\end{figure}

We also mention that the energy of the vortex becomes a nonzero constant
as $\alpha$ goes to infinity. Actually, we found that the dependence of the energy on
 $\alpha$ or the filling factor, $\pi/\alpha$,
is very small. The color-charge density of the condensate is very small at small filling factors, but
the value of the field $\langle\phi_a\rangle=v$ is never small. The energy of the soliton is governed by $\langle\phi_a\rangle$ so that
the energy never becomes small even as the filling factor becomes small.
In this respect,
we cannot find any instabilities of the QHS at small filling factors
$\pi/\alpha$. But as we expect, the state might be unstable 
at such a small filling factor because Wigner crystal is energetically 
more stable than the QHS at such small color-charge
density of gluons. We remember that the number difference of 
positively color-charged quarks and negatively charged quarks is
also small in such a case. Thus, for example, the excessive negatively charged quarks
form Wigner crystal. Therefore, gluons with positive charges also form Wigner crystal, not QHS
for reducing color Coulomb energy; in the QHS the color charge
distribution of the gluons is uniform. 
Accordingly, it is natural to expect that the QHS at much small
filling factor is unstable.

\subsection{effects of the third spatial dimension} 
Up to now, we have considered the ground-state structure of gluons in two-spatial dimension. 
This is because the unstable modes are two-dimensional objects and
they may form a stable ground state with their condensation. 
The unstable modes, in general, depend on all of the coordinates in three
dimensions; $\phi(k_3<\sqrt{g{\cal B}})\sim
\exp(ik_2x_2+ik_3x_3-iE(k_3)t)\exp(-(x_1-k_2\ell_B^2)^2/2\ell_B^2)$
with $E(k_3)=\sqrt{-g{\cal B}+k_3^2}$.
But, among them the modes with the largest amplitude
as $t\to \infty$, $\phi(k_3=0)$, depend only on $x_1$ and $x_2$.
They are two-dimensional objects and 
form the stable ground state, namely, the QHS of gluons 
as we have shown. In the derivation of the QHS
we have used Chern-Simons gauge theory, which can be used only
in two-dimensional space. In this way we have fully used two-dimensionality 
of the problem. We may wonder whether or not unstable modes with small, but nonvanishing
$k_3$ ( $\ll\sqrt{g{\cal B}}$ ) contribute to the ground state structure. 
We have a symmetry of the rotation around the magnetic field and 
of the translation along it. Thus, it is natural to expect that
the ground state should be uniform in $x_3$ direction. Then,
the modes should not be important, otherwise their contributions make
the ground state nonuniform in the direction.
Therefore, it is reasonable to expect that three-dimensional effects on the ground state change
our main result; the stable QHS of gluons is realized in
the ferromagnetic state ( Savvidy vacuum ).

We also found the gap energy, $\Delta $, above the ground-state energy
based on the two-dimensional theory. The effect of $x_3$ direction
is simply that the corresponding mode propagating in the direction
gains the energy, $\sqrt{\Delta^2+k_3^2}$. This is because the relativistic covariance in the direction 
still remains at least in the limit of infinitely large quark matter ( $\ell \to \infty$ ).
We may also wonder whether or not the gap-less mode with $E=|k_3|$
exists. In order to see it, we may assume that the fluctuation $\delta \phi_a$
does not have dependence on any spatial coordinates.
We found that there is no such 
solution; the condition of $\partial_3 \delta \phi_3=0$ has been taken 
into account explicitly in our treatment. 
 Therefore, the QHS we have found is really the stable gapped state of
the unstable gluons.

\section{color ferromagnetism vs.~color superconductivity} \label{cf}
Until now, quarks do not play any roles for the realization of
the ferromagnetic phase except for supplying color
charges for the condensate of the gluons. But, quarks play important roles
for the realization of the phase. Here, we briefly discuss 
their roles.

In general, the energy density of the quarks in the magnetic field is smaller than
that of the quarks without the magnetic field. ( This
fact is favorable to the ferromagnetic state. ) 
The fact is easily understood intuitively in the case of strong
magnetic field.
When the magnetic field is sufficiently strong, all of the quarks occupy
the lowest Landau level; their energy is given by $\sqrt{m_{\rm q}^2+k_3^2}$
where $k_3$ denotes the momentum parallel to the magnetic field 
of the quarks. 
Hence, the energy density of the quarks in the strong magnetic field is much lower than  
that of the quarks without the magnetic field. On the other hand, 
for sufficiently large number density of the quarks, equivalently,
for sufficiently weak magnetic field, the quarks occupy
much higher Landau levels. 
Eventually, both energy densities ( with and without $g{\cal B}$ ) approach to each other
in the limit of
$\rho \to \infty$. 

Actually, we can show that the energy density, $E_{\rm quark}(g{\cal B},\,\rho)$, of the quarks in the magnetic field is lower 
than $E_{\rm quark}(g{\cal B}=0,\,\rho)$ 
for any strength of the magnetic field $g{\cal B}$.
Thus, both energy densities approach each other very rapidly as $\rho \to \infty $;
$(E_{\rm quark}(g{\cal B},\,\rho)-E_{\rm quark}(g{\cal B}=0,\,\rho))/E_{\rm quark}(g{\cal B}=0,\,\rho) \leq O(\rho^{-4})$.

As has been pointed out, the energy density, $E_{\rm super}(\rho)$,  of the superconducting state of the quarks is also 
lower than
$E_{\rm quark}(g{\cal B}=0,\,\rho)$ due to the condensation of the quark's Cooper pairs. 
That is, only the quarks in  
the vicinity of the Fermi
surface whose width may be given by a gap energy $\Delta$,
gain energy $\Delta$ by making Cooper pairs. The decrease of the 
energy density is given such as $\sim \Delta^2\, \varepsilon_{\rm f}^2$ 
with the Fermi energy $\varepsilon_{\rm f}$; $\varepsilon_{\rm f} \to \rho^{1/3}$ as $\rho \to \infty$.
Normalizing it by $E_{\rm quark}(g{\cal B}=0,\rho)$, we find 
$\Delta^2\, \varepsilon_{\rm f}^2/E_{\rm quark}(g{\cal B}=0,\rho)\propto \Delta^2\,\varepsilon_{\rm f}^{-2} \to \rho^{-2/3}$.
Thus, the decrease of the energy in the BCS state 
is slower than that of the energy in the ferromagnetic state
when the number density of the quarks increases.
Therefore, when 
the number density of the quarks is sufficiently large, 
the color superconducting phase is energetically more favored\cite{color,ferro} 
than the ferromagnetic phase. 
We have also shown that
the energy decrease, $|E_{\rm quark}(g{\cal B},\,\rho)-E_{\rm quark}(g{\cal B}=0,\,\rho)|$,
of the quark state with the magnetic field is larger than the
energy decrease, $|E_{\rm super}(\rho)-E_{\rm quark}(g{\cal B}=0,\,\rho)|$,
of the superconducting state
when the number density of the quarks is much small. 
Thus, the ferromagnetic phase is realized.

In this way, the quarks play the role of realizing the superconducting phase
for sufficiently large chemical potential, while the
gluons play the main role of realizing the ferromagnetic phase
for the small chemical potential. 
The
quarks also
play the role of realizing the ferromagnetic phase
with chemical potential smaller than the one in the above.

\section{observational implication} \label{obs}

The QHS of gluons is realized in quark matter.
The matter can be produced by heavy ion collisions.
The matter produced in the collisions initially has high temperature
so that it is in the phase of the quark-gluon plasma. After that,
it gradually loses its energy and then enters into the phase of
the color ferromagnetic state with the QHS of gluons if
the value of the chemical potential is appropriate.
How do we detect whether or not the matter is in the phase?
We cannot observe the color magnetic field, which is confined in the matter.
But we show that the matter in the phase possesses 
a large observable magnetic moment, in other words, it
produces strong observable magnetic fields outside of the matter.
The point is that the difference between the number of positively color-charged
quarks and that of negatively charged quarks generates a rotation of the quark
matter of the ferromagnetic phase as a whole. The difference is a
result of the realization of the QHS.
When the quark matter 
is not electrically neutral, the rotation generates a magnetic 
moment. Suppose that the quark matter is composed of 
up and down quarks, and that the number difference between
positively and negatively color-charged quarks is identical in 
each flavor. We consider the QHS with $\nu=1/2$.
Then, the density difference, $\rho_f^{(-)}-\rho_f^{(+)}$, of 
each flavor is
given by the color-charge density of gluons in the QHS;
$\Delta\rho=\rho_f^{(-)}-\rho_f^{(+)}=g{\cal B}/8\pi \ell$.
A quark with electromagnetic charge $e_{\rm q}=2/3\,(\mbox{or}-1/3)\times e$ generates a 
magnetic moment ( $= \partial E(g{\cal B}+e_{\rm q}B)/\partial
B|_{B=0}$ ), which is
of the order of $e_{\rm q}/(2\sqrt{g{\cal B}})$ in the strong magnetic
field $\sqrt{g{\cal B}}>m_{\rm q}$. Hence, the magnetization of the quark
matter is given by

\begin{equation}
(2/3-1/3)e\frac{\Delta\rho}{2\sqrt{g{\cal B}}}=\frac{e\sqrt{g{\cal B}}}
{48\pi\ell}\sim (10\,\mbox{MeV})^2
\frac{\sqrt{g{\cal B}}}{800\,\mbox{MeV}}\frac{3\,\mbox{fm}}{\ell}
\simeq 1.4\times 10^{15} \,\mbox{Gauss}
\end{equation}
where we have assumed that the strength of the color-magnetic field is $\sqrt{g{\cal B}}=800$ MeV and 
the size of the quark matter is $\ell=3$ fm. We have taken the value of $g{\cal B}$ as a reference point based on 
the vacuum fluctuation of $<(g{\cal B})^2>\sim (800\mbox{MeV})^2$.   
The observation of this strong magnetic field can be an evidence of
the presence of the ferromagnetic phase in the quark matter.

\section{discussion} \label{dis}

In this paper we have discussed only the case of the SU(2) gauge theory. Similar
results hold even in the SU(3) gauge theory although possible structures
of QHSs are much richer in SU(3) case than in SU(2) case because of
the presence of more unstable modes\cite{ferro}. A particular point in
the SU(3) gauge theory is the presence of a phase with coexistence 
of color ferromagnetism and color superconductivity
( so-called 2SC )
at large chemical potential. This is because the direction of the
magnetic field in the color space is normal to the direction
of the quark pair condensate. The details should be referred to our
paper\cite{ferro}.

Quark confinement ( hadron phase ) is caused mainly by gluon's dynamics, namely, the SU(3)
gauge theory. Ground state structure of QCD is determined by
analyzing non-perturbative dynamics of gluons in the phase. Especially, we need fully dynamical
treatment in the gauge theory for revealing the property of the confinement.
The quarks play no dominant roles for the confinement. For example,
in the large $N$ expansion of the SU($N$) gauge theory the confinement is realized at
$0$-th order of the expansion, in which the quark loops do not arise.
The contribution of the quarks appears in the higher order of the
expansion so that the effects of the quarks can be treated perturbatively
in the expansion. 

On the other hand, the dynamics of the quarks play an important role for
the color superconductivity in the region of large chemical potential of
the quark number. Gluons simply give perturbation, an 
attractive force in an appropriate 
channel of the quarks; it makes the Fermi gas of free quarks unstable and realizes the superconducting state
of the quarks. 

The color ferromagnetic phase is realized between the hadron phase and
the color superconducting phase when the chemical potential is varied.
Thus, it is natural to expect that both gluon and quark dynamics play important roles
for the phase. As we have explained, indeed, the gluon dynamics 
plays 
the main role in leading to the stable
ferromagnetic phase along with quantum Hall state of gluons 
when the quark matter is present.
In such a case, the quark dynamics plays a role of choosing the ferromagnetic phase when
the chemical potential is small. On the other hand at large
chemical potential the quark dynamics plays a role of 
leading to and of choosing the superconducting phase.
In other words, the gluon dynamics plays a main role in realizing the
ferromagnetic state at small chemical potential, while the quark
dynamics plays a main role in realizing
the color superconducting
state at large chemical potential.

In this way, in QCD the carrier of the main role 
for determining various phases of quark matter changes 
from the gluons to the quarks when we increase the chemical potential of
the quark number.

\hspace*{1cm}

Two of the authors ( A. I. and M. O.) 
express thanks to the member of theory 
group in KEK for their hospitality.



\begin{thebibliography}{99}
\bibitem{color}K. Rajagopal and F. Wilczek, {\tt hep-ph/0011333}.
\bibitem{monopole}S. Mandelstam, Phys. Lett. {\bf 53B}, 476 (1975).\\
G. 't Hooft, Nucl. Phys. {\bf B190}, 455 (1981).
\bibitem{abelian}Z.F. Ezawa and A. Iwazaki, Phys. Rev. {\bf D25}, 2681
  (1982).
\bibitem{ferro}A. Iwazaki and O. Morimatsu, Phys. Lett. {\bf B571}, 61
 (2003); A. Iwazaki, O Morimatsu, T. Nishikawa and M Ohtani,
 Phys. Lett {\bf B579}, 347 (2004).
\bibitem{nielsen}N.K. Nielsen and P. Olesen, 
Nucl. Phys. {\bf B144}, 376 (1978);
Phys. Lett. {\bf 79B}, 304 (1978).
\bibitem{savidy}G.K. Savvidy, Phys. Lett. {\bf 71B}, 133 (1977).\\
H. Pagels, Lecture at Coral Gables, Florida, 1978.
\bibitem{qh} {\it The Quantum Hall Effect, 2nd Ed.}, edited by R.E. Prange
  and S.M. Girvan ( Springer-Verlag, New York, 1990 ).
\bibitem{iwa}Z.F. Ezawa, M. Hotta and A. Iwazaki, 
Phys. Rev. {\bf B46}, 7765 (1992);
Z.F. Ezawa and A. Iwazaki, J. Phys. Soc. Jpn. {\bf 61}, 4133 (1990). 
\bibitem{qhs}K.von Klitzing et al., Phys. Rev. Lett. {\bf 45}, 494 (1980). 
\bibitem{tsui}D.C. Tsui, H.L. Stormer and A.C. Gossard,
  Phys. Rev. Lett. {\bf 48}, 1559 (1982).
\bibitem{laughlin}R.B. Laughlin, Phys. Rev. Lett. {\bf 50}, 1359 (1983).
\bibitem{zhang}S.C. Zhang, H. Hanson and S. Kilvelson          
Phys. Rev. Lett. {\bf 62}, 82 (1989).
\bibitem{iwa2}F.Z. Ezawa, M. Hotta and A. Iwazaki, Mod. Phys. Lett. {\bf B6},
  737 (1992). 
\bibitem{fruct}L. Saminadayar, D.C. Glattli, Y. Jin and B. Efienne,
  Phys. Rev. Lett. {\bf 79}, 2526 (1997).\\
R. de Picciotto, M. Reznikov, M. Heiblum, V. Umansky, G. Bunin and
D. Mahalu, Nature 389, {\bf 162}, (1997).
\bibitem{jo} Z.F. Ezawa and A. Iwazaki, Phys. Rev. {\bf B47}, 7295 (1993);
I.B. Spielman, et al. Phys. Rev. Lett. {\bf 84}, 5808 (2000); M.M. Fogler
and F. Wilczek, Phys. Rev. Lett. {\bf 86}, 1833 (2001).
\bibitem{nn}J. Ambj\o{}rn, N.K. Nielsen and P. Olesen, Nucl. Phys. {\bf B152},
  75 (1979).\\
H.B. Nielsen and M. Ninomiya, Nucl. Phys. {\bf B156}, 1
  (1979).\\
H. B. Nielsen and P. Olesen, Nucl. Phys. {\bf B160}, 330 (1979).
\bibitem{nakajima} T. Nakajima and M. Ueda, Phys. Rev. Lett. {\bf 91} 140401 (2003).
\bibitem{seme}G.W. Semenoff, Phys. Rev. Lett. {\bf 61}, 516 (1988); 
G.W. Semenoff and P. Sodano, Nucl. Phys. {\bf B328}, 753 (1989).
\bibitem{iwa3}Z.F. Ezawa and A. Iwazaki, Phys. Rev. {\bf B47}, 7295 (1993).
\bibitem{spa}H.B. Nielsen and P. Olesen, Nucl. Phys. {\bf B160}, 380 (1979).
\end{thebibliography}
\end{document}